\documentclass{aa}
\usepackage[varg]{txfonts}
\usepackage[colorlinks=true,     linkcolor=blue, citecolor=blue, filecolor=blue, urlcolor=blue]{hyperref}

\usepackage{natbib}
\bibpunct{(}{)}{;}{a}{}{,} 

\begin{document}

\title{Identification of a Single Plasma Parcel during a Parker Solar Probe-Solar Orbiter radial alignment}

\author{Etienne Berriot\inst{1}
    \and Pascal D\'emoulin\inst{1}$^,$\inst{2}
    \and Olga Alexandrova\inst{1}
    \and Arnaud Zaslavsky\inst{1}
    \and Milan Maksimovic\inst{1}}

\institute{LESIA, Observatoire de Paris, Universit\'e PSL, CNRS, Sorbonne Universit\'e, Universit\'e Paris Cit\'e, 5 place Jules Janssen, 92195 Meudon, France
  \and Laboratoire Cogitamus, rue Descartes, 75005 Paris, France
  }

\abstract{
Configurations where two spacecraft, such as Parker Solar Probe (PSP) and Solar Orbiter (SolO), are radially aligned provide opportunities to study the evolution of a single solar wind parcel during so called plasma line-ups.
The most critical part of such studies is arguably the identification of what can be considered a same plasma crossing both spacecraft.
We present here a method that allowed us to find what we believe to be the same plasma parcel passing through PSP ($\sim 0.075$~au)  and SolO ($\sim 0.9$~au) after their radial alignment the 29/04/2021.
We started by modeling the plasma propagation in order to get a first estimation of the plasma line-up intervals.
The identification of the same density structure (with a crossing duration $\sim 1.5$~h) passing through the two spacecraft allows to precise and confirm this estimation.
Our main finding is how stable the density structure is, remaining well recognizable from PSP to SolO despite its $\sim 137$ hours journey in the inner heliosphere.
We moreover found that the studied slow solar wind plasma parcel undergone a significant acceleration (from $\sim 200$ to $\sim 300$ km/s) during its propagation.
}


\maketitle


\nolinenumbers 

\section{Introduction} \label{sec:intro}

The solar corona’s temperature is too high to remain in hydrostatic equilibrium.
This leads to an expansion of the Sun's atmosphere in the interplanetary medium, creating a supersonic and super-Alfv\'enic outflow of plasma we call the solar wind \citep{Parker_1958}.

The solar wind is often separated into two categories depending on its speed, the fast and slow winds.
The fast solar wind is thought to arise from coronal holes \citep{Zirker_1977_fast_wind_coronal_holes, McComas_1998_Ulysses_winds_latitudes_minimum_activity} with speeds ranging from $\sim$500 km/s to $\sim$800 km/s.
The slow solar wind has lower temperatures and higher densities than the fast wind, with speeds ranging from $\sim$150 km/s to $\sim$500 km/s, where the lowest speeds are only observed when close to the Sun \citep{Sanchez-Diaz_2016_VSSW, Maksimovic_2020_electrons, Dakeyo_2022_isopoly}.
The processes giving it birth as well as its exact origins are however still disputed  \citep[see][and references therein]{Abbo_2016_slow_solar_wind_review,Rouillard_2021_solar_wind_review}.
There are also still many pending questions regarding the radial evolution of the solar wind and interplanetary structures it carries within the heliosphere.

Helios 1 \& 2 allowed the study of radial evolution of what can be considered the same plasma passing through both spacecraft when they were radially aligned.
This method was first employed by \citet{Schwenn_line-ups_1981_b,Schwenn_line-ups_1981_a}, giving it the name "plasma line-up".
One of the interval they found has been thoroughly studied in \citet{Schwartz_Marsch_1983_radial_evolution} when Helios 1 and Helios 2 respectively were situated at $\sim$0.51~au and $\sim 0.72$~au from the Sun.
Their study focuses on the radial evolution of the energy budget and adiabatic invariants \citep{Chew_1956_CGL} in what was identified to be the same plasma parcel inside a fast solar wind stream.
The authors also discuss different hypotheses and difficulties linked with the mapping and plasma parcel identification.
Their identification has been done considering a constant and radial propagation speed for the plasma parcel.
However, any acceleration, or non-radial flow can change the time intervals to consider and are therefore potential sources of uncertainties.

Moreover, the term "plasma line-up" and "plasma parcel" are somewhat ill-defined, as discussed in \citet{Schwartz_Marsch_1983_radial_evolution}.
For example, the particles populations of the solar wind tend to propagate with different velocities.
The protons often exhibit a beam in their velocity distribution function, the electrons have a thermal speed much higher than their bulk speed, and alpha particles usually have a different bulk speed's than the protons \citep[see][and references therein]{Marsch_2012_VDF_Helios}.
Furthermore, halo electrons are probably governed by non-local scattering mechanisms \citep{Zaslavsky_2024}.
We however prefer to keep the general terminology "plasma parcel/line-up" throughout this study.

The recently launched Parker Solar Probe \citep[PSP,][]{Fox_2016_Parker_Solar_Probe_science_goals} and Solar Orbiter \citep[SolO,][]{Muller_2020_Solar_Orbiter_science_goals} missions 
are great new opportunities for such line-up studies.
Both of them are orbiting in the inner heliosphere simultaneously, allowing combined observations and measurements between the two spacecraft \citep{Velli_2020_PSP_SolO}.
Conjoined observations with other spacecraft such as STEREO-A or BepiColombo are also possible \citep{Hadid_2021_Bepi_cruise_phase_synergies}.

\begin{figure*}[t!] 
\centering
   \includegraphics[width=17cm]{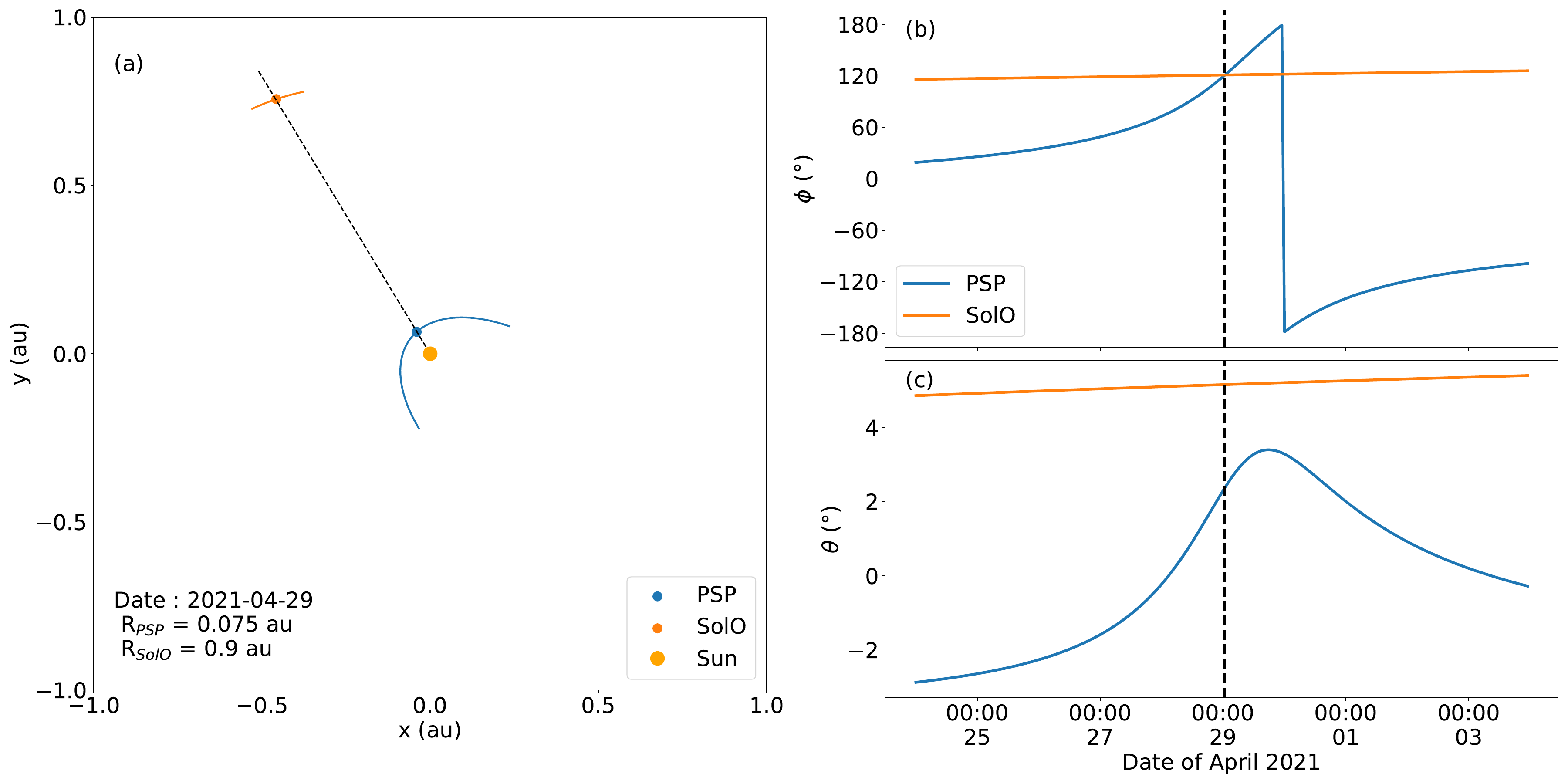}
\caption{
On panel (a) are PSP (blue) and SolO (orange) positions the 29/04/2021 (dots) and trajectories (lines) within the ecliptic plane between the 24/04/2021 and the 04/05/2021 as seen in an inertial reference frame centered on the Sun. The radial line coming from the Sun and passing through PSP and SolO for the spacecraft coalignment time $t_{ 0}$ is indicated by a black dashed line.
Panels (b) and (c) respectively show the longitude $\phi$, and latitude $\theta$ of PSP (blue) and SolO (orange) as functions of time for the same interval as in panel (a). We have indicated the spacecraft coalignment time $t_{ 0} =$ 00:45~UT on the 29/04/2021 as a vertical black dashed line on both panels.
}
\label{fig:line-up-config_fused}
\end{figure*}  

Indeed, there have been some recent plasma line-up studies, with PSP (0.1 au) and SolO (1 au) in \citet{Telloni_2021}, and with PSP (0.17 au) and BepiColombo (0.6 au) in \citet{Alberti_2022}.
The purpose of these studies was the radial evolution of statistical properties of magnetic turbulence. 
\citet{Telloni_2021} first estimated the intervals for the plasma line-up assuming a constant speed of 315~km.s$^{-1}$ for the solar wind (as measured at PSP).
They then assessed the plasma correspondence by calculating the Pearson correlation coefficient between PSP and SolO magnetic field's magnitude measurements at time intervals corresponding to different propagation speeds ([250, 350] km.s$^{-1}$) to take into account some eventual acceleration.
A similar method for the plasma propagation has been employed in \citet{Alberti_2022}.
The authors also based their identification on a cross-correlation method between the two spacecraft magnetic field's magnitude, with sliding windows for both PSP and BepiColombo and calculation of the mutual information coefficient \citep{Shannon_1948_a, Shannon_1948_b, Cover_Thomas_2005_information_theory_book}.
The identifications in \citet{Telloni_2021} and \citet{Alberti_2022} suggest nearly zero acceleration of the solar wind during the propagation, a result not verified due to the absence of solar wind velocity measurement at the outer spacecraft in both studies.
This is quite surprising as several statistical studies reported a non-negligible acceleration of the slow solar wind in the inner heliosphere \citep{Schwenn_line-ups_1981_b, Sanchez-Diaz_2016_VSSW, Maksimovic_2020_electrons, Dakeyo_2022_isopoly}.

The results of plasma line-up studies are highly dependent on the time intervals taken as being the same parcel of solar wind.
It is therefore crucial to be able to unambiguously identify what can be considered the same plasma on both spacecraft.

In this paper, we present a new approach that allowed us to identify the same plasma parcel passing through two radially aligned spacecraft. We study here a radial alignment between PSP (at $\sim 0.075$~au) and SolO (at $\sim 0.9$~au) on the 29/04/2021. The identification of the same plasma is done through several steps.
We first present an overview of the spacecraft configuration in Section~\ref{sec:line-up-config}.
Then, in Section~\ref{sec:propag_method}, we describe our modeling of the solar wind's propagation from the inner to outer spacecraft to estimate the time intervals corresponding to the plasma line-up.
Using this estimation, we identify the same density structure passing through both PSP and SolO in Section~\ref{sec:plasma_identification}.
We finally summarize our results and conclude in Section~\ref{sec:conclusion}.

\section{Data and Line-up Configuration} \label{sec:line-up-config}

The goal of this study is to identify the same plasma parcel at two different distances from the Sun with PSP and SolO.
In order to do this, we had to take advantage of a configuration where the two spacecraft are quasi radially aligned as shown in Figure~\ref{fig:line-up-config_fused}(a), where we consider the positions of PSP and SolO on the 29/04/2021 around 01:00 \footnote{
The ephemerides were obtained from SPICE kernels \citep[][\url{https://naif.jpl.nasa.gov/naif/}]{Acton_1996_NAIF}, for both PSP (\url{https://spdf.gsfc.nasa.gov/pub/data/psp/ephemeris/spice/}) and SolO (\url{https://doi.org/10.5270/esa-kt1577e}).
}.
The plasma parcel is first crossed by PSP, then propagates outward and is eventually crossed by SolO after some propagation time $\tau$.

For this configuration, PSP and SolO were respectively situated at approximately 0.075~au and 0.9~au from the Sun.
We show in Figure~\ref{fig:line-up-config_fused}(b) and (c) the longitude ($\phi$) and latitude ($\theta$) of the two spacecraft around their radial alignment, defined as the time when they are both at the same longitude as indicated by the vertical black dashed lines.
We will call this time $t_{ 0}$. It corresponds to 
   \begin{equation} \label{eq:t0}
   t_0 = \text{29/04/2021 00:45 UTC}
   \end{equation}
and is used as the reference time for the rest of this study.
We therefore define
\begin{equation}
    t = t_{UTC} - t_0
\end{equation}
where $t_{UTC}$ is the Coordinated Universal Time.

We remark in Figure~\ref{fig:line-up-config_fused}(b) that PSP's longitude is varying  much more than SolO's one. This is because PSP is orbiting much faster than SolO as it is about 12 times closer to the Sun.
Indeed, at $t_0$, $\omega_{PSP} \simeq 1.25 \times 10^{-5}$ rad/s and $\omega_{SolO} \simeq 1.95 \times 10^{-7}$ rad/s, so $\omega_{PSP} / \omega_{SolO} \sim 64$, with $\omega_{PSP}$ and $\omega_{SolO}$ the angular speeds of PSP and SolO respectively.

We also point out that there is a latitude difference of $\Delta \theta \sim 3$° at $t_{0}$ when the two spacecraft are coaligned in longitude.
At SolO's distance ($\sim 0.9$ au), this difference corresponds to a length $l_{\Delta \theta} \simeq 7 \times 10^6$~km or $0.05$ au. This $l_{\Delta \theta}$ imposes a lower bound on the scale of the plasma parcel we can expect to observe on both spacecraft.

For this study\footnote{
The used data are publicly available at (\url{https://spdf.gsfc.nasa.gov/pub/data/psp}), for PSP’s SWEAP and FIELDS (\url{https://doi.org/10.48322/0yy0-ba92}) measurements, and at \url{http://soar.esac.esa.int/soar/}, for SolO’s SWA (\url{https://doi.org/10.5270/esa-ahypgn6}) and MAG (\url{https://doi.org/10.5270/esa-ux7y320}) measurements.
}
, we analyzed protons densities $N_p$, and bulk velocities $\mathbf{V}_{p}$ as well as the magnetic field $B$ measured by PSP and SolO.
For the protons, on PSP those are obtained with measurements of the SPAN-ion instrument \citep{Livi_2021_SPAN-ion}, part of the SWEAP \citep{Kasper_SWEAP_Instrument_PSP} suite, while on SolO, we are using measurements of the Proton and Alpha particle Sensor \citep[PAS,][]{Owen_2020_SWA_SolO}. 
Magnetic field measurements are from the FIELDS instrument \citep{Bale_FIELDS_Instrument_PSP} on PSP and the MAG instrument on SolO \citep{Horbury_2020_MAG_SolO}.

\begin{figure}[t!] 
\resizebox{\hsize}{!}{\includegraphics{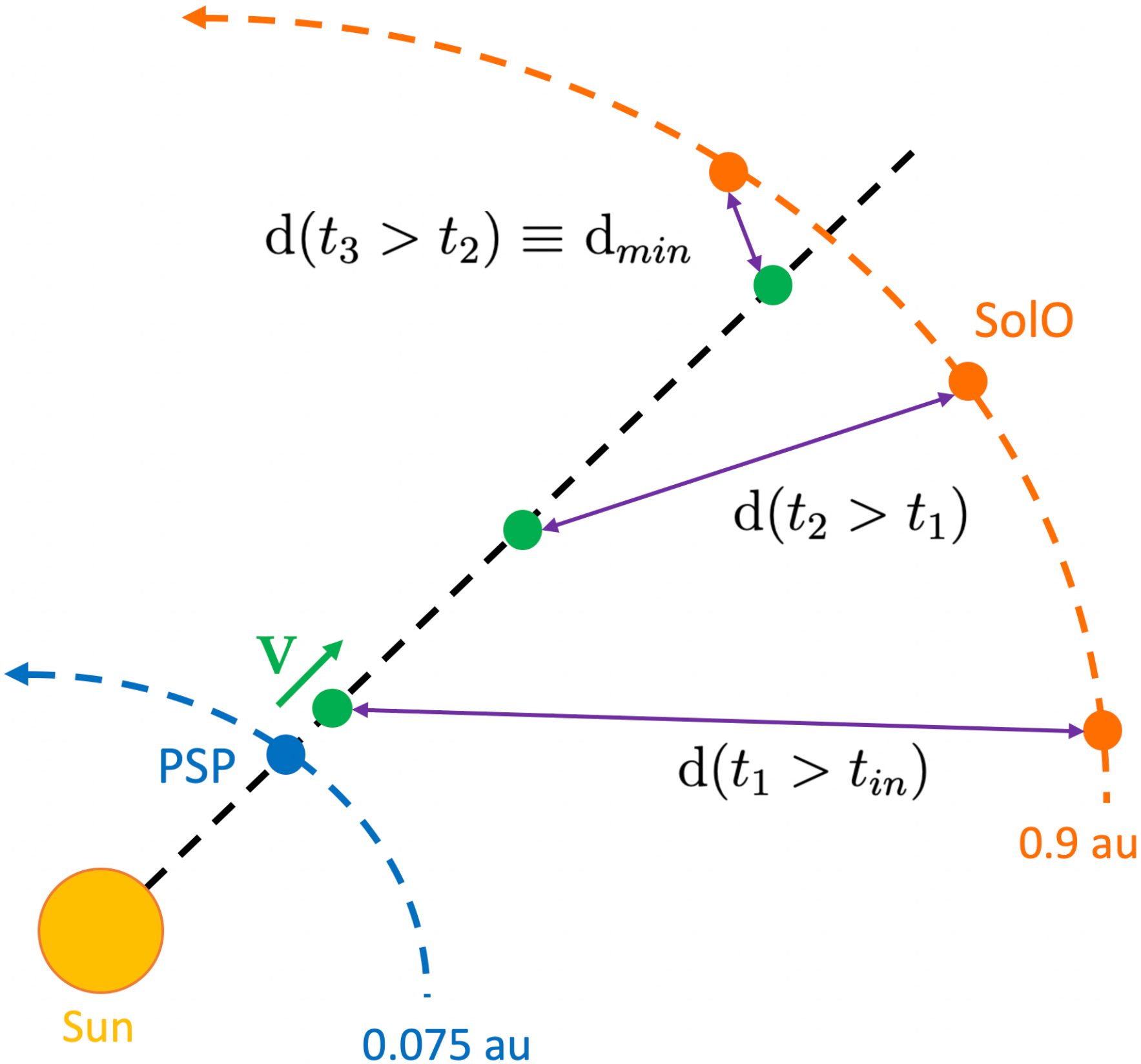}}
\caption{Simplified and not to scale schematic of the propagation method for a purely radial plasma speed. PSP's position at a considered time $t_{in}$ is indicated by a blue dot. We represented with green dots, and at subsequent times ($t_1, t_2, t_3$), the position of the plasma parcel passing by PSP at $t_{in}$ and propagating outward with $\mathbf{V}$ (green vector).
We also represented the position of SolO (orange dots) and the distance between the parcel and SolO (purple double arrows) at the same times.
The simplified trajectories of PSP and SolO are shown by blue and orange dashed lines respectively.
The plasma parcel trajectory is shown by the black dashed line.}
\label{fig:propag-method-schematic}
\end{figure}  

\section{Ballistic Propagation model} \label{sec:propag_method}

In order to identify the time intervals corresponding to a plasma line-up, one first needs to model the propagation of the plasma parcel from the inner spacecraft to the outer one.
There are several ways of doing this modeling, we propose the following method.

We begin with the configuration shown in Figure~\ref{fig:line-up-config_fused}, for which the two spacecraft are quasi-aligned.
The positions of PSP and SolO as functions of time $t$ will be noted $\mathbf{R}_{PSP}(t)$ and $\mathbf{R}_{SolO}(t)$ respectively.
We then define the position of the plasma parcel, $\mathbf{R}(t, t_{in})$, at every moment $t$ following its crossing of the inner spacecraft at a time $t_{in}$: 
   \begin{equation} \label{eq:Rparcel}
   \mathbf{R}(t, t_{in}) = \mathbf{R}_{ in}(t_{in}) + \int_{t_{in}}^{t} \mathbf{V}(t',t_{in}) dt' 
   \end{equation}
where $\mathbf{V}(t',t_{in})$ is the plasma propagation velocity, which can have any profile, and by definition $\mathbf{R}_{ in}(t_{in}) = \mathbf{R}_{PSP}(t=t_{in})$.


\begin{figure}[t!] 
\resizebox{\hsize}{!}{\includegraphics{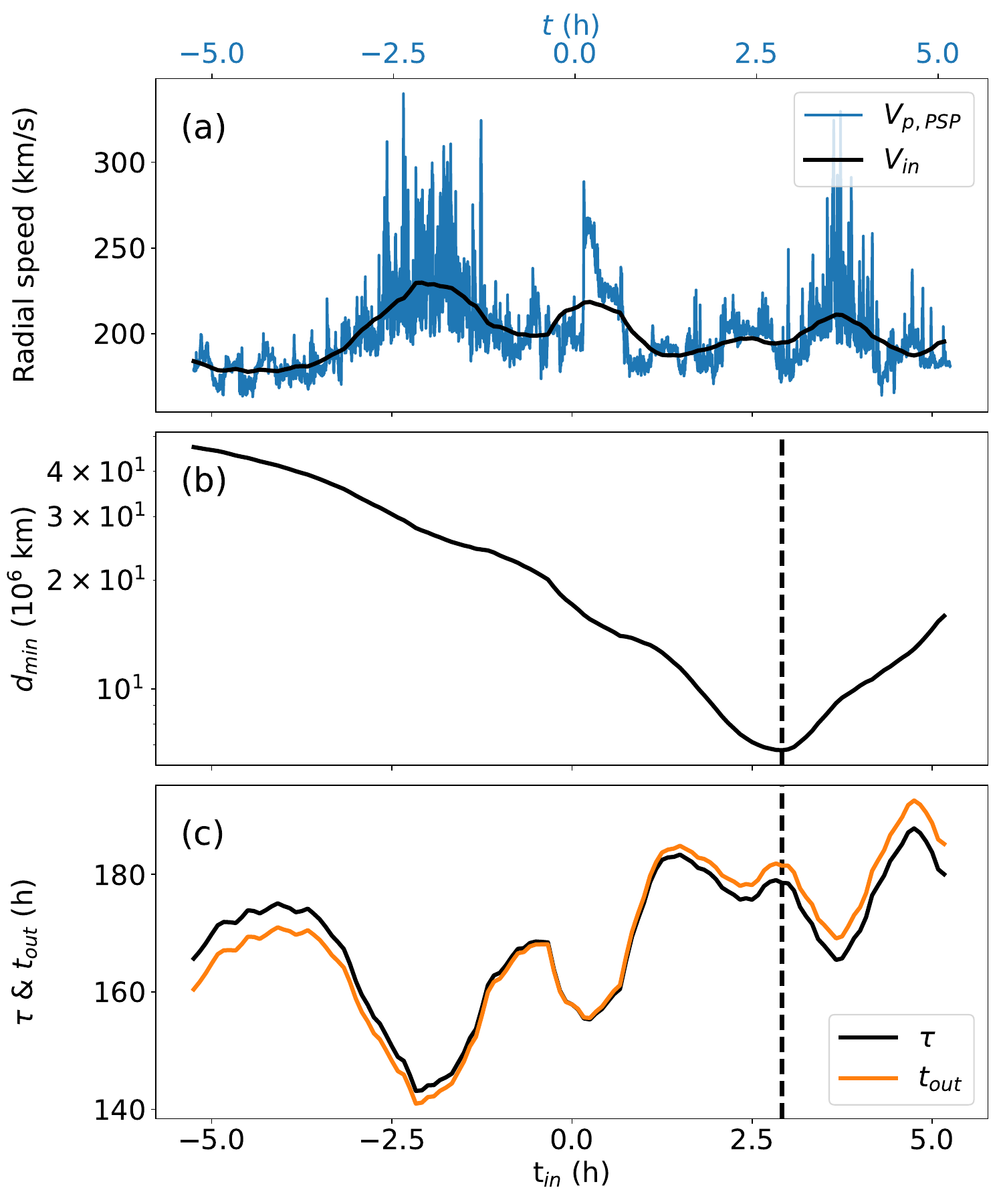}}
\caption{
Outcome of the propagation method for a radial and constant velocity $\mathbf{V} = (V_R = V_{in}, 0, 0)$ in RTN coordinates.
Panel (a) shows the radial proton speed recorded by PSP ($V_{p,PSP}(t)$ in blue) and $V_{in} (t_{in})$ (in black) calculated as averages of $V_{p, PSP}$ over $\Delta t = 1$~h centered on each considered $t_{in}$.
Panels (b) and (c) respectively show the minimum distance $d_{min}$ between the plasma parcel and SolO, and the corresponding propagation times $\tau$ and $t_{out}$, all functions of $t_{in}$.
The vertical lines in panels (b) and (c) correspond to the $t_{in}$ for which $d_{min}$ is minimum (called $d_{MIN}$).
All the time origins are set at $t_0$, corresponding to the radial alignment time of the spacecraft (Equation (\ref{eq:t0}))
}
\label{fig:propag-method-outcome}
\end{figure}  

\begin{figure}[t!] 
\resizebox{\hsize}{!}{\includegraphics{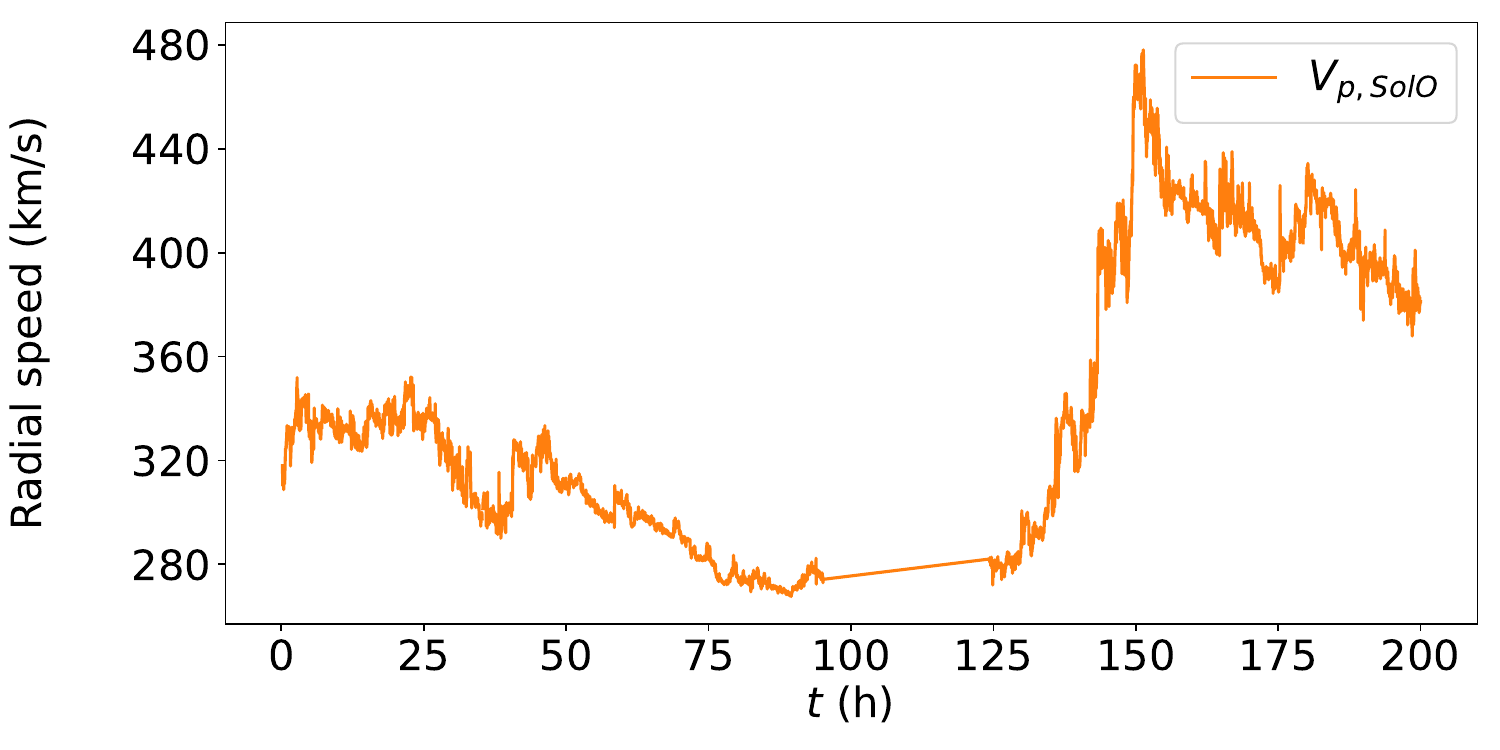}}
\caption{
Solar wind radial speed recorded by SolO ($V_{p, SolO} (t)$), for a time interval corresponding to $0 \leq t \leq 200$~h. The time origin is set at $t_0$ (Equation (\ref{eq:t0})).
}
\label{fig:SolO-speed-LS}
\end{figure}  

We calculate the distance between the plasma parcel and the outer spacecraft as
  \begin{equation}
  d(t, t_{in}) = |\mathbf{R}_{SolO}(t) - \mathbf{R}(t, t_{in})| \label{eq:d(t,tin)}
  \end{equation}
The process is schematized in Figure~\ref{fig:propag-method-schematic}.
After some travel time $\tau(t_{in})$, this distance passes by a minimum $d_{ min} (t_{in})$ at a time $t = t_{out} (t_{in})$.
We define the position of the plasma parcel after the travel as $\mathbf{R}_{ out}(t_{in}) \equiv \mathbf{R}(t=t_{out}, t_{in})$.
We next iterate the above computation for different $t_{in}$.
Each considered time $t_{in}$ is therefore linked to a time $t_{out}(t_{in})$ through the relation:
  \begin{equation}   \label{eq:propag-time}
  t_{out}(t_{in}) = t_{in} + \tau(t_{in})
  \end{equation}
This allows to describe, for every considered solar wind parcel, when the plasma crossed by the inner spacecraft (at $t_{in}$) gets the closest to the outer one (at $t_{out}(t_{in})$), as well as the corresponding distance $d_{ min}(t_{in})$ and propagation time $\tau(t_{in})$.

\subsection{Propagation Method with Constant Velocity} 
\label{sec:velocity_constante}

We computed the propagation method described above with the following settings.
We considered a set of starting times $t_{in}$, spaced by 5 min, each of which defines the starting position of one plasma parcel.
For every $t_{in}$, we define the associated parcel's propagation velocity as $\mathbf{V}_{in} \equiv \langle \mathbf{V}_{p, PSP} \rangle$, where $\langle \mathbf{V}_{p, PSP} \rangle$ is the average protons velocity measured by PSP over a time interval $\Delta t$ centered on $t_{in}$.
We chose ${\Delta t} = 1$~h as to get a more relevant estimation of the plasma propagation's speed by averaging velocity fluctuations related to the turbulent cascade.
Finally, the plasma parcel positions are computed for every $t > t_{in}$, with a 1 min resolution on $t$, and that for each $t_{in}$.

We show in Figure~\ref{fig:propag-method-outcome} the results of this propagation method for the line-up configuration of Figure~\ref{fig:line-up-config_fused}.
Figure~\ref{fig:propag-method-outcome}(a) shows the radial protons bulk speed measured by PSP $V_{p, PSP}(t)$ with a blue curve, and $V_{in} (t_{in})$ with a black curve.
Figure~\ref{fig:propag-method-outcome}(b) gives the estimated distance between the plasma parcel and the outer spacecraft after the propagation, $d_{ min}(t_{in})$, while on Figure~\ref{fig:propag-method-outcome}(c) are the corresponding propagation time $\tau (t_{in})$ and time at the outer spacecraft $t_{out} (t_{in})$.

The closest approach is well defined: $d_{ min}$ has a clear minimum $d_{ MIN} = \text{min}(d_{ min}) \simeq 7 \times 10^6$~km, indicated by the vertical dashed line.
Therefore the time intervals to look at the inner and outer spacecraft should respectively be around the associated $t_{in} \simeq 2.9$~h and $t_{out} \simeq 180$~h.
We obtained $d_{ MIN} \simeq 7 \times 10^6$~km, which is close to $l_{\Delta \theta}$ (due to the spacecraft latitude difference around $t_{ 0}$) estimated in the end of Section~\ref{sec:line-up-config}.
This is expected since the corresponding $t_{in} (\simeq 2.9$~h) and $\tau (\simeq 180$~h) are respectively much lower than the orbiting period of PSP and SolO, therefore the spacecraft latitude difference $\Delta \theta$ does not change significantly from its value at $t=t_0$.
Due to the large radial difference between the spacecraft, the speed variations observed at PSP (Figure \ref{fig:propag-method-outcome}(a)), and therefore on $V_{in}$ make the propagation time $\tau (t_{in})$ vary a lot (145~h $\lesssim \tau \lesssim $ 185~h) in the range of $t_{in}$ considered, see Figure~\ref{fig:propag-method-outcome}(c).

We next look at the solar wind recorded by SolO for a very large range of times to see how well the hypothesis of constant speed is verified.
We thus show on Figure~\ref{fig:SolO-speed-LS} the proton radial speed recorded by SolO for $0 \leq t \leq 200$~h.
Unfortunately, there is no SWA data available from the $\sim$05/02/2021 23:50 to the $\sim$05/04/2021 05:00, which translates to $90 \leq t \leq 125$~h.
Even though, no matter the considered $t \in [0, 200]$~h, the observed proton's speed at SolO is higher than at PSP.
This is consistent with results of precedent studies that reported an acceleration of the slow wind in the inner heliosphere \citep{Sanchez-Diaz_2016_VSSW, Maksimovic_2020_electrons, Dakeyo_2022_isopoly}.
We therefore have to take this acceleration into account in the propagation method.

\subsection{Propagation Method with Constant Acceleration} \label{sec:acceleration_constante}

Using only the data at both spacecraft, the \textit{in situ} measured speeds can only define an average acceleration during the travel time. In the absence of other data, the simplest solution is to consider a constant acceleration constrained by the protons velocities measured by PSP and SolO.
However, this acceleration $\mathbf{a}$ is not directly accessible from the two set of measurements as it first requires to link $t_{in}$ and $t_{out} (t_{in})$ through the minimization of $d(t, t_{in})$ (Equation (\ref{eq:d(t,tin)})) for each $t_{in}$.  This is done below by scanning a range of $\mathbf{a}$, constrained by the observations.
In order to simplify the notations, we will mostly omit explicit mentions of the dependencies to $t_{in}$ in the rest of this paper.

We first consider the plasma propagation with an arbitrary constant acceleration $\mathbf{a}$.
For each $t_{in}$, the positions and speeds of the plasma parcel at every time $t > t_{in}$ following the crossing of the inner spacecraft are given as
  \begin{eqnarray}
  \mathbf{R}(t) =& \mathbf{R}_{ in} + (t - t_{in}) \mathbf{V}_{in} + \frac{(t - t_{in})^2}{2} \, \mathbf{a} \nonumber\\ 
  \mathbf{V}(t) =& \mathbf{V}_{in} + (t - t_{in}) \, \mathbf{a}          \label{eq:V_acceleration}
  \end{eqnarray}
After the propagation this model provides
  \begin{eqnarray}
  \mathbf{R}_{ out} = & \mathbf{R}_{ in} + \tau \mathbf{V}_{in} + \frac{\tau^2}{2} \mathbf{a} \label{eq:position-parcel-tout} \\
  \mathbf{V}_{out} = & \mathbf{V}_{in} + \tau \, \mathbf{a}  \label{eq:speed-parcel-tout}
  \end{eqnarray}
with $\tau$ the propagation time, $\mathbf{R}_{out} = \mathbf{R}(t=t_{out})$, and $\mathbf{V}_{out} = \mathbf{V}(t=t_{out})$.
We use Equation (\ref{eq:speed-parcel-tout}) to write the acceleration as
  \begin{equation} \label{eq:acceleration-tau} 
  \mathbf{a} = \frac{\mathbf{V}_{out} - \mathbf{V}_{in}}{\tau}
  \end{equation}
Replacing this in Equation (\ref{eq:position-parcel-tout}), we get
\begin{equation} 
  \tau = \frac{2 \, ||\mathbf{R}_{ out} - \mathbf{R}_{ in}|| }{|| \mathbf{V}_{out} + \mathbf{V}_{in} ||} \nonumber \,.\label{eq:propagtime-vectorielle}
\end{equation}
Combining the two last equations, we finally can express the acceleration as
  \begin{equation} \label{eq:acceleration-vectorielle-finale}
  \mathbf{a} = \frac{|| \mathbf{V}_{out} + \mathbf{V}_{in} ||}{2\, ||\mathbf{R}_{ out} - \mathbf{R}_{ in}|| }\left( \mathbf{V}_{out} - \mathbf{V}_{in}  \right) 
  \end{equation}
If we indeed cross the same plasma on both spacecraft, then after the propagation we should have $\mathbf{R}_{ out} \approx \mathbf{R}_{SolO}(t=t_{out})$ and $\mathbf{V}_{out} \approx\mathbf{V}_{p, SolO}(t=t_{out})$.
$\mathbf{V}_{in}$ can be estimated using $\mathbf{V}_{p, PSP}(t)$ around $t_{in}$ and $\mathbf{R}_{ in} = \mathbf{R}_{PSP} (t=t_{in})$.
The acceleration written in Equation~(\ref{eq:acceleration-vectorielle-finale}) is therefore a function of the observed quantities only, provided that $t_{in}$ and $t_{out}$ are defined.
However, it does not incorporate the association between $t_{in}$ and $t_{out}$ established by minimizing $d(t, t_{in})$ of Equation~(\ref{eq:d(t,tin)}).

\begin{figure}[t!] 
\resizebox{\hsize}{!}{\includegraphics{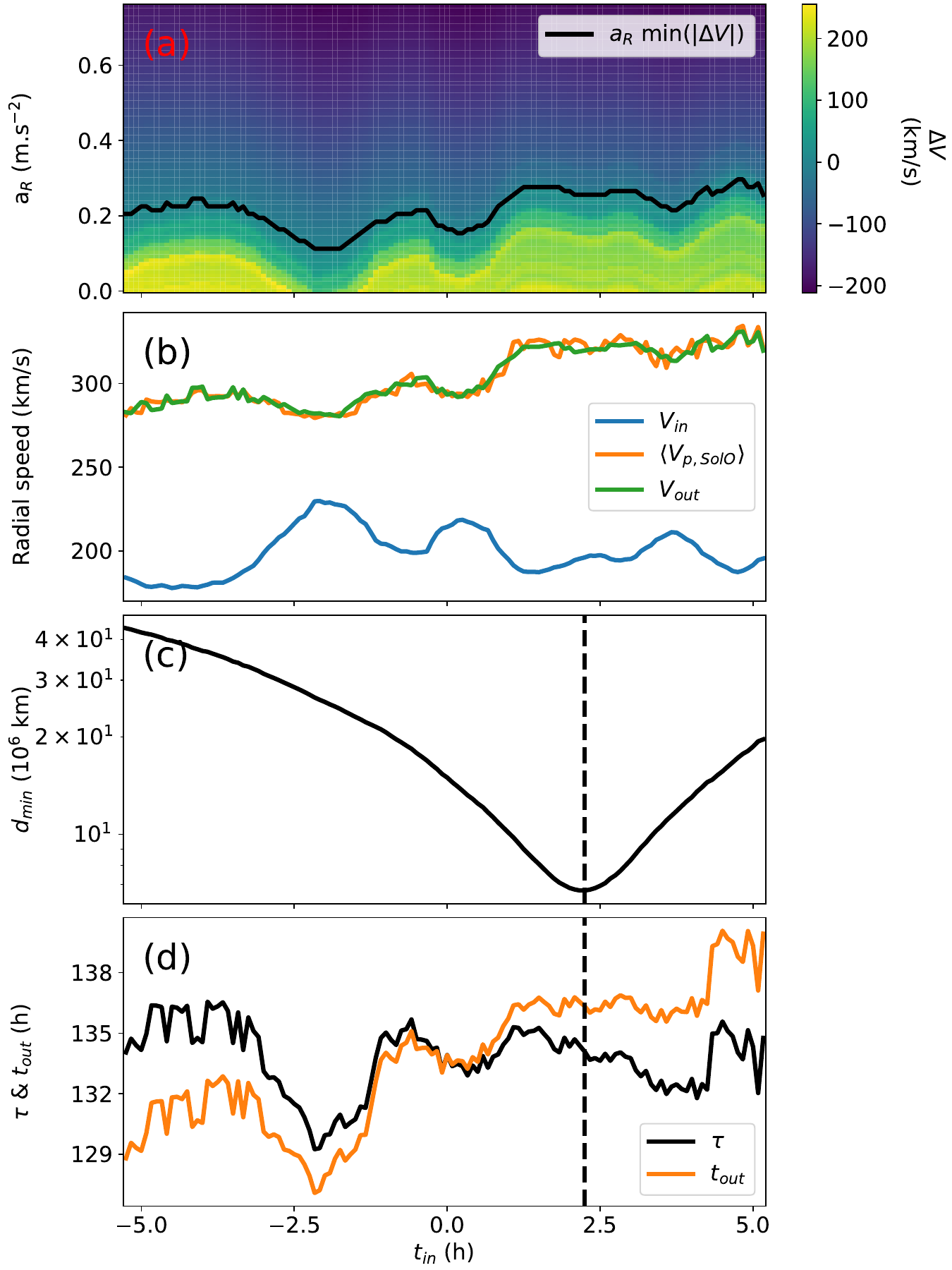}}
\caption{
Propagation method outcomes using a constant acceleration $a$ constrained by the observed plasma speeds at the inner and outer spacecraft.
All quantities are shown as functions of $t_{in}$.
The colors on panel (a) are values of $\Delta V= \langle V_{p, SolO} \rangle - V_{out}$ and the black curve is the radial acceleration for the minimum of $|\Delta V|$.
The method's outcomes shown in panels (b), (c) and (d) are all corresponding to propagations considering this acceleration.
On panel (b) are, in blue $V_{in}$ calculated as in Figure \ref{fig:propag-method-outcome}, in orange $\langle V_{p, SolO} \rangle$ defined as the proton bulk speed measured by SolO and averaged for $\Delta t =$1~h centered on $t_{out}$, in green $V_{out}$, which is the theoretical speed given by Equation (\ref{eq:speed-parcel-tout}).
The corresponding $d_{ min}$, $\tau$ and $t_{out}$ are shown in panels (c) and (d) respectively.
}
\label{fig:outcomes_acceleration}
\end{figure}  

The results are expected to be close to the spacecraft alignment, so $R_{in}$ and $R_{out}$ are weakly dependent of the precise $t_{in}$ and $t_{out}$ values.
We therefore calculate $||\mathbf{R}_{out} - \mathbf{R}_{in}||$ using the position of PSP around $t_{in}$ and the position of SolO around $t_{out}$ (both associated to $d_{MIN}$) estimated using a constant velocity (Figure~\ref{fig:propag-method-outcome}).
This approximation is a posteriori justified by the fact that the $||\mathbf{R}_{ out} - \mathbf{R}_{ in}||$ estimated using a constant velocity and the one estimated with a constant acceleration only differ by $\sim 1 \%$.
The acceleration given by Equation~(\ref{eq:acceleration-vectorielle-finale}) is indeed mostly changing with $t_{in}$ due to variations of $\mathbf{V}_{in}$ and $\mathbf{V}_{out}$.

As previously, we consider a purely radial plasma propagation and $V_{in} = \langle V_{p, PSP} \rangle$ averaged over $\Delta t = 1$~h around $t_{in}$.
The acceleration and speeds are noted as scalars representing the radial component of the vectors.
Since the radial acceleration cannot be derived directly from observations, we define a maximum acceleration $a_{ max}$ as
  \begin{equation}  \label{eq:a_max}
  a_{ max} = \frac{V_{out, max}^2 - V_{in, min}^2}{2 \, ||\mathbf{R}_{out} - \mathbf{R}_{in}||}
  \end{equation}
where we fix $V_{out, max} = 480$~km/s and $V_{in, min} = 180$~km/s close to the maximum and minimum of the radial plasma speeds respectively measured by SolO and PSP during the considered time periods.
Then, for every $t_{in}$, we compute 75 plasma propagations, all with a different constant acceleration. For this, we considered accelerations values uniformly spaced between 0 and $a_{max}$. By minimizing $d(t, t_{in})$ for each $a$ value we obtain the parameters ($t_{out}$, $\tau$, $d_{MIN}$) as functions of $t_{in}$ and $a$.

We next need to define the $a$ value which is the most compatible with the observed velocity at SolO for each $t_{in}$ value.
The above model provides $V_{out} = V(t = t_{out})$ using Equation (\ref{eq:V_acceleration}) projected on the radial axis.
Using SolO measurements, we also define (as for PSP) a proton bulk speed $\langle V_{p, SolO} \rangle$ averaged over $\Delta t = 1$~h around each obtained $t=t_{out}$. 
We note that because of the data gap at SolO (straight line in Figure~\ref{fig:propag-method-outcome}(d)), we considered the missing data as interpolated values between the nearest available measurements.
We therefore define
  \begin{equation}  \label{eq:DeltaV}
  \Delta V = \langle V_{p, SolO} \rangle - V_{out}
  \end{equation}
Finally, for each $t_{in}$ we select the acceleration $a$ that minimizes $|\Delta V|$ so that the propagation model with a constant acceleration is most consistent with the observations at both spacecraft.

We show the results of the above procedure in Figure~\ref{fig:outcomes_acceleration}.
On panel (a), the colors indicate values of $\Delta V$ with the black curve being the acceleration associated to the minimum of $|\Delta V|$ at each $t_{in}$.
Panel (b) shows the corresponding $V_{in}$ in blue, $\langle V_{p, SolO} \rangle$ in orange and $V_{out}$ in green. 
The minimum of $|\Delta V| $ is therefore the difference between the green and orange lines. Its standard deviation is 
$\simeq 2$ km/s.
This small value, an order of magnitude lower than the fluctuations of $V_{in}$ and $\langle V_{p, SolO} \rangle$, shows that the acceleration bins are numerous enough.

The minimum distance $d_{ min}$, obtained for the minimum of $|\Delta V|$, is shown in Figure~\ref{fig:outcomes_acceleration}(c).
As in Figure~\ref{fig:propag-method-outcome}, the vertical dashed line indicates the value $t_{in}$  obtained for the minimum distance $d_{ MIN}$. Here, $t_{in} = 2.25$~h which is smaller by only $\sim 40$~min as compared to the result obtained above with a constant velocity.
Considering a purely radial propagation speed and neglecting 3D effects, the plasma line-up implies that the longitudes $\phi$ of the inner and outer spacecraft have to be the same at $t_{in}$ and $t_{out}$, respectively. Using the longitude of the spacecraft line-up as the longitude origin, this condition approximately writes
  \begin{equation}
  \omega_{out} \, t_{out} = \omega_{in} \, t_{in}  
  \end{equation}
where $\omega_{out}$ and $\omega_{in}$ are the longitudinal angular velocities of the spacecraft around the Sun (in an inertial reference frame).
This implies that a variation of $t_{out}$, say by $\delta t_{out}$ due to a different transit time $\tau$, has an effect $\delta t_{in} = (\omega_{out}/\omega_{in}) \delta t_{out}$ on $t_{in}$.
So, a high angular velocities ratio $\omega_{out}/\omega_{in}$ between the two spacecraft, due to an important difference in their distances from the Sun, leads to low $\delta t_{in}$.
With the present PSP and SolO configuration the ratio $\omega_{out}/\omega_{in} \approx 1/64$.
This explains why in the studied case, the $t_{in}$ associated with the plasma line-up is weakly dependent on the plasma velocity profile.

Figure~\ref{fig:outcomes_acceleration}(d) shows $\tau$ and $t_{out}$.
These $t_{out}$ values are fortunately not in the data gap shown in Figure~\ref{fig:SolO-speed-LS}, so the resulting outcomes are directly linked to SolO observations.  $\tau$ and $t_{out}$ fluctuate much less than with a constant speed hypothesis (Figure~\ref{fig:propag-method-outcome}(c)).
More precisely, for a 10 hours time interval of $t_{in}$, the $t_{out}$ variations are in a time interval of 15~h for a constant acceleration, as compared to 50~h in the case of a constant velocity.
Hence, the constant acceleration model (constrained by both spacecraft data) provides a more reliable estimation of the time at SolO for an approximate plasma line-up.
The constant acceleration model implies also a significantly lower propagation time $\tau$, by almost 50~h as compared to the constant velocity model (as expected due to the observed increase of plasma velocity).

\begin{figure*}[t!] 
\centering
   \includegraphics[width=17cm]{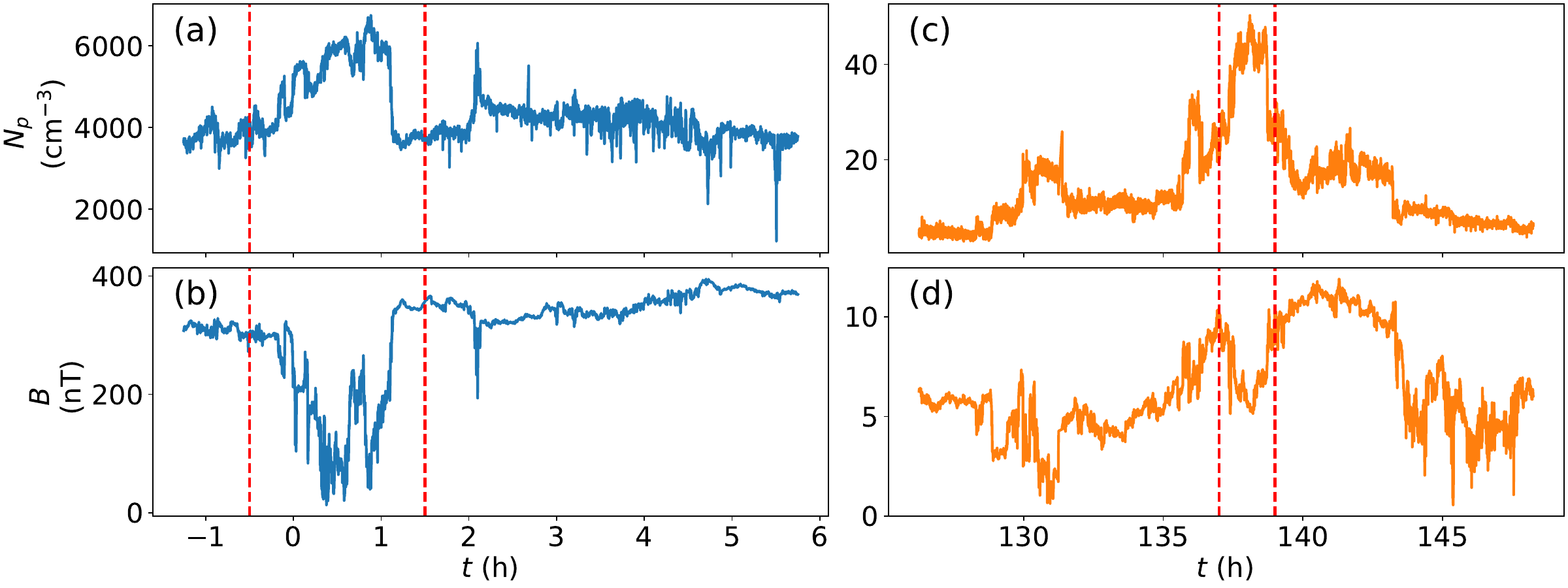}
\caption{
Proton density $N_p$ (panels (a) and (c)) and magnetic field's magnitude $B$ (panels (b) and (d)) measured by PSP (blue) around $t_{in}$ and Solar Orbiter (orange) around $t_{out}$ when the minimum distance $d_{ MIN}$ is achieved.
}
\label{fig:raw_data_cc}
\end{figure*}  

Figure~\ref{fig:outcomes_acceleration}(c) shows that $d_{MIN} = 7 \times 10^6$~km (the minimum value of $d_{min}$), is well defined.
This $d_{MIN}$  is close to the one obtained with a constant velocity (Figure~\ref{fig:propag-method-outcome}) and the distance estimated in Section~\ref{sec:line-up-config} due to the spacecraft latitude difference around the spacecraft alignment ($l_{\Delta \theta} \sim 7 \times 10^6$~km).
The fact that $d_{MIN} \simeq l_{\Delta \theta}$ confirms that, when considering a purely radial propagation velocity, this minimum is mainly constrained by the spacecraft orbits and not by the plasma dynamics. 
Moreover, the $d_{ min}$ curve with constant accelerations, Figure~\ref{fig:outcomes_acceleration}(c), is smoother than with constant velocities (Figure~\ref{fig:propag-method-outcome}(b)). This is a consequence of a narrower range of $t_{out}$ as shown in Figure~\ref{fig:outcomes_acceleration}(d).

\section{Same Plasma Identification} \label{sec:plasma_identification}

\subsection{Data selection}\label{sec:data_selection_cross-correl}

The propagation model provides a first estimation of the time intervals corresponding to the plasma line-up.
This estimation however has to be made more precise and confirmed.
To do so, we propose to search for a same structure passing through both spacecraft.
Once found, this structure can then be used as a marker to unambiguously define the same plasma.

On Figure~\ref{fig:raw_data_cc} are the proton density $N_p$ and magnetic field's magnitude $B$ measurements taken by PSP (panels (a, b)) and SolO (panels (c, d)) around the times corresponding to $d_{ MIN}$, so $t_{in} = 2.25$~h and $t_{out} \sim 135$~h as determined above, see Figure \ref{fig:outcomes_acceleration}(d).
We therefore chose to consider $t \in 2.25 \pm 4$~h for PSP, and $t \in 135 \pm 10~$h for SolO to take into account uncertainties linked to the propagation model.
We remark on PSP measurements the presence of a density enhancement and simultaneous anti-correlated magnetic field's depletion for $t \in [-0.5, 1.5]~$h (indicated by red vertical dashed lines).
We note that a visual comparison with SolO's data already allows the identification of a very similar structure around $t\in [137,139]~$h, also indicated with two red vertical dashed lines.

The proton density is indeed typically a useful parameter to look as it often exhibits well identifiable spatial structures.
Density enhancements have been reported in the solar wind (especially the slow one), either using remote-sensing instruments \citep[e.g.][]{Sheeley_1997_blobs, Rouillard_2010_transients_a, Viall_2015_PDS} or \textit{in situ} measurements \citep[e.g.][]{Viall_2008_PDS_lengths_in-situ, Rouillard_2010_transients_b, Stansby_2018_density_structures}.
These structures, that are thought to emerge near the tip of coronal helmet streamers, are also generally believed to be well conserved during their propagation in the heliosphere and simply carried with the surrounding solar wind \citep{Kepko_2016_PDS_composition_and_slow_wind_source, DiMatteo_2019_PDS_Helios, Kepko_2024_PDS_formed_at_Sun}, "like leaves in the wind" as put by \citet{Sheeley_1997_blobs}.
However, we notice that, in opposition to the leaves, in the solar wind these plasma structures are themselves part of the medium.

\begin{figure}[t!] 
\resizebox{\hsize}{!}{\includegraphics{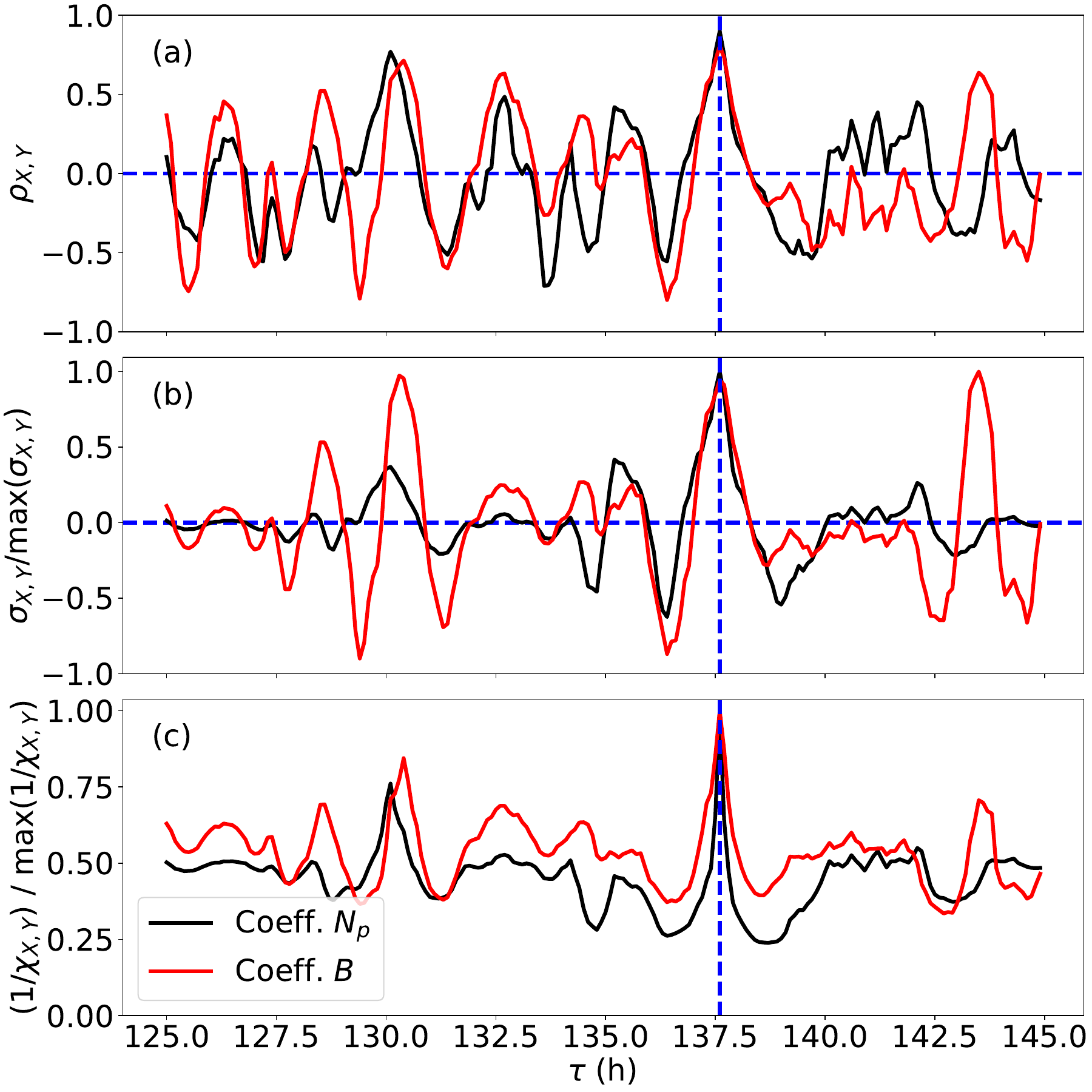}}
\caption{
Cross-correlation for measurements of proton density $N_p$ (black curves) and magnetic field magnitude $B$ (red curves) for PSP and SolO using different measures. Panel (a) shows the Pearson correlation coefficient defined by Equation~(\ref{eq:rho_XY}), panel (b) the covariance, and panel (c) $1/\chi_{X,Y}$ defined by Equation~(\ref{eq:chi}).
The abscissa $\tau$ is the travel time between PSP and SolO. The time interval considered at PSP is the one shown between dashed lines on Figure~\ref{fig:raw_data_cc}(a,c). The same time duration, 2~h, is used for the moving time window at SolO.}
\label{fig:cross-correl-structures}
\end{figure}  

Moreover mesoscale structures, such as density enhancements, typical radial size at 1 au are $\sim 5 \times 10^3 - 10^7$ km \citep[][and references therein]{Viall_2021_mesoscale_structures}.
Then, the larger density structures have sizes comparable to the minimum distance $d_{min}$ ($\in [1, 1.5] \times 10^7$~km for $t_{in} \in [-0.5, 1.5]$~h) evaluated for a propagation with constant acceleration.
This implies that, since SolO is situated at $\sim 0.9$~au in the studied case, one of those density structures could be large enough to eventually be crossed by both spacecraft.
We note that $d_{min}$ is most probably lower due to non-radial propagation effects (Appendix \ref{sec:Non-radial-propagation}).
Furthermore, a direct comparison between the radial sizes of such structure (determined as $V_{p,R} \times \delta t$, with $\delta t$ the duration of the structure) and $d_{min}$ is only pertinent for structures with similar radial and ortho-radial extensions.
Due to the solar wind's nearly spherical expansion, we however expect to have structures elongated in the latitudinal and longitudinal directions at SolO.

\subsection{Association with Cross-Correlations} \label{sec:cross-correl}

The goal is here to see if the structure identified at PSP (between two red vertical dashed lines in Figure~\ref{fig:raw_data_cc} (a,b)) has also passed through SolO.
A common way to quantitatively assess the correspondence between the measurements at two spacecraft is through the use of a cross-correlation method.
Therefore, using three different coefficients, we computed the cross-correlation between PSP measurements for $t \in [-0.5, 1.5]~$h, and intervals of same temporal lengths ($T = 2$~h) at SolO for time shifts $125$~h~$\leq \tau \leq 145$~h.
We chose a time step of 0.1~h between each $\tau$, defining the intervals at SolO used for the cross-correlation.
We show the results in Figure~\ref{fig:cross-correl-structures}, where we computed these coefficients for both the proton density $N_p$ (black curves) and the magnetic field amplitude $B$ (red curves).

Figure \ref{fig:cross-correl-structures}(a) shows the cross-correlation function based on the standard Pearson correlation coefficient $\rho_{X,Y}$, defined as
    \begin{equation}   \label{eq:rho_XY}
   \rho_{X,Y}(\tau) = \frac{\langle \delta X(t) \, \delta Y(t+\tau) \rangle}
    {\sqrt{\langle \delta X(t)^2 \rangle} \sqrt{\langle \delta Y(t + \tau)^2 \rangle}} 
    \end{equation}
with
\begin{eqnarray}
    \delta X(t) &=& X(t) - \langle X(t) \rangle \nonumber \\
    \delta Y(t + \tau) &=& Y(t + \tau) - \langle Y(t + \tau) \rangle 
\end{eqnarray}
for $X(t)$ and $Y(t + \tau)$ the same physical parameter recorded by PSP and SolO respectively.
The brackets $\langle ... \rangle$ denote averages over $T$ (=2~h here).
Due to its normalization, this coefficient is not affected by the amplitudes of $X$ and $Y$, and therefore tends to give high correlation between structures of different amplitudes.
This explains why we observe numerous high values peaks in Figure \ref{fig:cross-correl-structures}(a).
 
On Figure~\ref{fig:cross-correl-structures}(b), is the cross-correlation function based on the covariance $\sigma_{X,Y}$ defined as
\begin{equation} \label{eq:covariance}
    \sigma_{X,Y}(\tau) = \langle \delta X(t) \, \delta Y(t+\tau) \rangle 
\end{equation}
and which we normalized by its maximum over all the tested intervals (max($\sigma_{X,Y}$)). That way, due to their large variance, the biggest structures will therefore tend to naturally give higher values for this coefficient.

On Figure~\ref{fig:cross-correl-structures}(c) is the cross-correlation function based on the inverse of the chi-square coefficient $1/ \chi_{X,Y}$, also normalized by its maximum over all the tested intervals, with $\chi_{X,Y}$ defined as :
    \begin{equation}  \label{eq:chi}
    \chi_{X,Y}(\tau) =   \sqrt{\langle \left( \delta X_c(t) - \delta Y_c(t+\tau) \right)^2 \rangle}  \,.
    \end{equation}
$\chi_{X,Y}$ is named from its similarity with the statistical chi-square and should be minimum when the signals are the most similar.
We therefore chose to show $1/\chi_{X,Y}$ to more easily compare it with cross-correlation functions based on other coefficients.
We use the typical radial variation summarized with a power law of solar radial distance to define the normalised quantities $\delta X_c(t) = \delta X(t) \, ( R_{X}/R_0)^\varepsilon$ and $\delta Y_c(t+\tau) = \delta Y(t + \tau) \, (R_{Y}/R_0)^\varepsilon$, where $R_{X}$ \& $R_{Y}$ are the distance between the spacecraft and the Sun, and $R_0=1$~au is the distance taken for normalisation.
We fixed $\varepsilon = 2$ for the density and $\varepsilon = 1.6$ for the magnetic field's magnitude in order to take into account the plasma's nearly spherical expansion.
Due to its spiral shape, the interplanetary magnetic field magnitude falls of less rapidly than $R^{-2}$. We therefore choose to correct $B$ by a factor $(R/R_0)^{1.6}$, in accordance with previous statistical studies using Helios 1 data \citep{Musmann_1997_IMF_radial_variations, Schwenn-Marsch_book_I_1990}.
This correction is needed as the solar wind expansion changes strongly the magnitude of the studied plasma parameters with solar distance. 
Note that, because of the subtraction $\delta X_c(t) - \delta Y_c(t+\tau)$, Equation (\ref{eq:chi}) results are sensitive to the choice of $\varepsilon$ (or any other normalization chosen for the signals).
However, $1/\chi_{XY}$ is an important quantity since it shows much less numerous peaks, with the main one being more outstanding and narrower than with the two other correlation coefficients
Finally, the three above coefficients are all linear and any non-linear evolution of the plasma will therefore not correctly be taken into account.
\begin{table}
\caption{Correlation coefficients between PSP and SolO measurements for the proton density $N_p$ and the magnetic field's magnitude $B$ at $\tau = 137.6$~h (see dashed line in Figure~\ref{fig:cross-correl-structures} and Section~\ref{sec:cross-correl}).
Due to the normalization, a value of 1 for the coefficients using $\sigma_{X,Y}$ and $1 / \chi_{X,Y}$ indicates an absolute maximum over the tested intervals.
\label{tab:correl-coeff}}
\centering
\begin{tabular}{clll}
\hline \hline
& \multicolumn3c{Correlation coefficients} \\
Physical Quantity & $\rho_{X,Y}$ & $\frac{\sigma_{X,Y}}{\text{max}(\sigma_{X,Y})}$ & $\frac{1/ \chi_{X,Y}}{\text{max}(1/ \chi_{X,Y})}$ \\
\hline
$Np$   & 0.90 & 1 & 1    \\
$B$    & 0.81 & 0.97 & 1    \\
\hline
\end{tabular}
\end{table}

All these cross-correlation functions show either absolute or local maxima for Np and B at $\tau = 137.6$~h (Figure~\ref{fig:cross-correl-structures}).
We show those values in Table \ref{tab:correl-coeff}.
The fact that all three coefficients have such high values, for both $N_p$ and $B$, at $\tau = 137.6$~h supports the correspondence between the two structures shown within the red vertical dashed lines in Figures \ref{fig:raw_data_cc} (a,b) and (c,d).

As we can see on Figure~\ref{fig:cross-correl-structures}, the correlation coefficients also exhibit local maxima for some other time shifts.
The most remarkable one being for $\tau \sim 130$~h.
Going back to Figure~\ref{fig:raw_data_cc}(c, d), we see that there are another density structure and associated magnetic field's depletion around $t = 130$~h.
The corresponding correlation values however are lower and a following visual inspection does not allow to find a clear correspondence between the structures.
We therefore disregard this density structure as being the same than the one observed at PSP.

\subsection{Justifications and limitations of the correlation method}

In the previous subsection, several parameters involved in the correlation estimation were fixed.
We justify below this choice and present a more general way of computing the different correlation coefficients.
We point out that, in general $X$ and $Y$ can be described as functions of several free parameters:
\begin{eqnarray}
    X =& X(t^*, T_X, \delta t_X) \nonumber \\
    Y =& Y(t^* + \tau, T_Y, \delta t_Y) \nonumber
\end{eqnarray}
with $t^*$ denoting the center of the time interval at the inner spacecraft, $\tau$ the time shift between $t^*$ and the center of the time interval at the outer spacecraft (propagation time).
$T_X$ and $T_Y$ are the lengths of the time intervals, and we also introduce $\delta t_X$ and $\delta t_Y$, the time resolution over which to resample the $X$ and $Y$ data sets respectively.
The evaluation of the correlation coefficient however requires that of $X$ and $Y$ have the same number of values $n$.
This adds a constraint:
\begin{equation*}
    \frac{T_X}{\delta t_X} = \frac{T_Y}{\delta t_Y} = n.
\end{equation*}
Therefore, any correlation coefficient $C_{X,Y}$ between two measurements $X$ and $Y$ should in general be computed as a function of all the free parameters:
\begin{equation*}
    C_{X,Y} = C_{X,Y} (t^*, \tau, T_X, T_Y, n).
\end{equation*}
In Section \ref{sec:cross-correl}, we fixed and constrained these parameters, using hypotheses discussed below, as to get physically relevant results.

As previously specified, we have set $t \in [-0.5, 1.5]~$h, fixing $t^* =0.5$~h and $T_X = 2$~h.
The goal was to select a prominent structure passing through one of the spacecraft, to then search and find it at the other spacecraft.
We selected the structure on PSP first because, for the plasma line-up, $t_{in}$ is better defined than the associated $t_{out}$, which is more dependent on the model selected to describe the plasma velocity (Section~\ref{sec:acceleration_constante}).

We also supposed that the selected structure had a global uniform evolution with a homogeneous acceleration during the propagation.
As we will see later (Figure \ref{fig:radial_stretching}, Section \ref{sec:Association}), this implies that the structure's temporal duration is similar at both spacecraft.
This is why the time window $T$ (= 2~h here) used for the cross-correlation is the same on both spacecraft $T_X = T_Y = T$.

Regarding the temporal resolution, we have $T_X = T_Y$, so $\delta t_X = \delta t_Y = \delta t$, and  we chose $\delta t = 20$~s.
We note that in our case the cross-correlation results were weakly dependent on $\delta t$ as long as $\delta t \ll T$.
Finally, the above constraints leave only $\tau$ as a free parameter.

Written in this form, the set of Equations (\ref{eq:rho_XY}, \ref{eq:covariance}, \ref{eq:chi}) are valid for the special case of $T_X = T_Y$.
Those Equations can however simply be written in their generalized form, for whatever signals $X$ and $Y$ respectively constituted of $n$ samples $X_i$ and $Y_i$, by replacing the bracket $\langle ... \rangle$ by $\frac{1}{n} \sum_i^n (...)$ and $X(t) \, \& \, Y(t + \tau)$ by $X_i \, \& \, Y_i$.

In summary, we have shown that the cross-correlation method can be misleading as high values for one (or several) coefficient do not necessarily imply that the same structure has been crossed by both spacecraft.
Therefore, such method has to be employed carefully, ideally on several (relevant) physical quantities, and using different coefficients.
Finally, although it can be helpful, the use of such method alone is not enough for a rigorous identification of the same plasma at two spacecraft.
The identification can only be confirmed through a physical analysis of the measurements.

\begin{figure}[t!] 
\resizebox{\hsize}{!}{\includegraphics{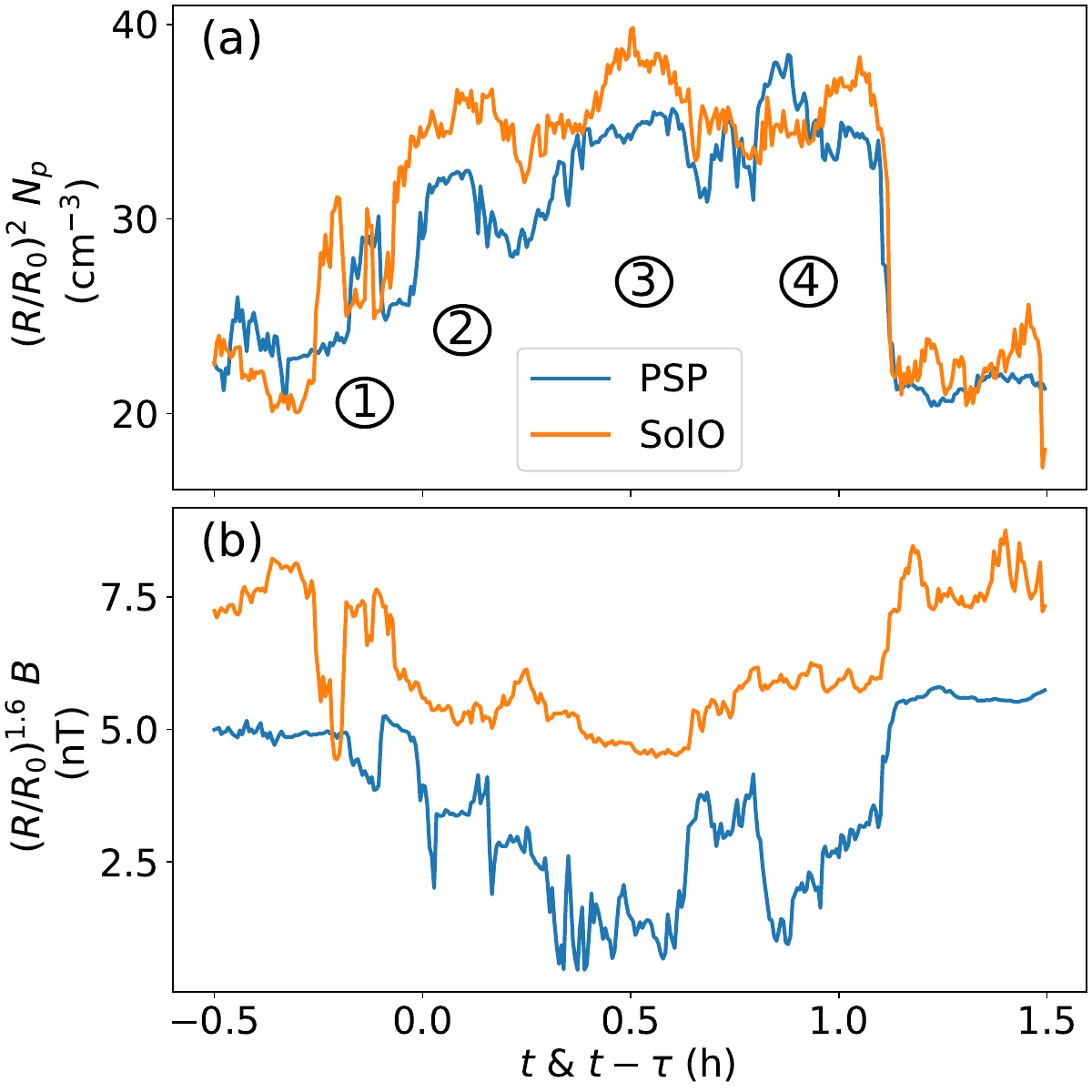}}
\caption{Proton density structure (a) and magnetic field magnitude (b), corrected by a factor $(R/R_0)^2$ and $(R/R_0)^{1.6}$ respectively, to take into account the expansion from PSP (blue) to SolO (orange).
Quantities are plotted as functions of $t$ for PSP and as functions of $t - \tau$ for SolO over a 2 h time interval (with $\tau= 137.6$~h as precised by several cross correlation methods, see Section~\ref{sec:cross-correl}).
An average over 20~s is applied to the data to better highlight the global behaviour of the density structures.
}
\label{fig:structure-brute}
\end{figure}  

\subsection{Local comparison of the structures}
\label{sec:Association}
 
We show in Figure~\ref{fig:structure-brute} the proton density and the magnetic field magnitude measurements of PSP and SolO taking into account the propagation time  $\tau= 137.6$ h, adjusted using a cross-correlation methods, as described in Section~\ref{sec:cross-correl}.
The measurements on PSP are plotted as functions of $t$ over a 2~h time interval, (between the two red vertical dashed lines in Figure \ref{fig:raw_data_cc}).
SolO measurements have been plotted as functions of $t - \tau$ as to get comparable signals, also over a 2 hour time interval. 
We correct $N_p$ by a factor $(R/R_0)^{2}$ and $B$ by a factor $(R/R_0)^{1.6}$, with $R$ the spacecraft distance to the Sun and $R_0 = 1$~au (as done in the definition of $\chi_{X,Y}$, see Section $\ref{sec:cross-correl}$).

\begin{figure}[t!] 
\resizebox{\hsize}{!}{\includegraphics{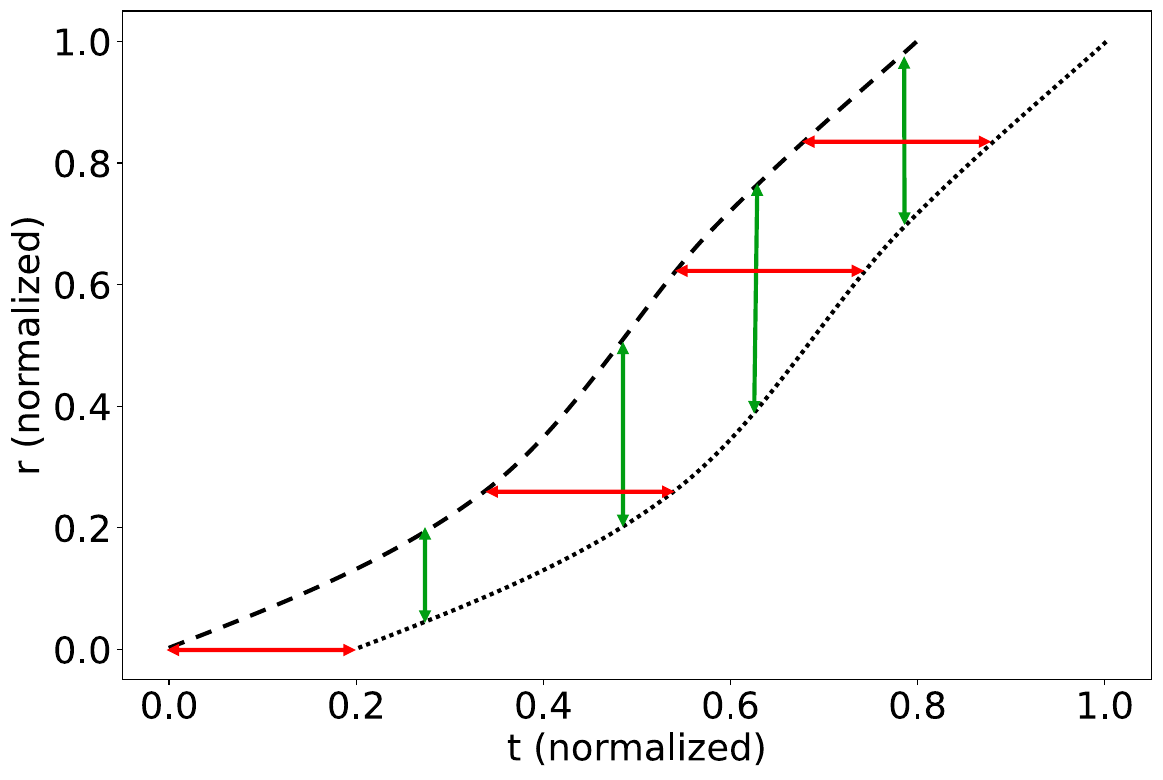}}
\caption{
Schematic illustration of a 1D structure propagation.  The structure is delimited by its rear and front boundaries (with dashed and dotted lines, respectively). These boundaries are supposed to have the same acceleration profile (so the same forces are acting on them).
We represent here a general case with arbitrary acceleration and deceleration.
Their normalized position (r) is shown as a function of normalized time (t).
The red horizontal arrows represent the time interval of the structure at a given radial position (as typical measurements provided by spacecraft).
Green vertical arrows represent the spatial length of the structure at a given time.}
\label{fig:radial_stretching}
\end{figure}  

We note that there is not only a global correspondence, as found in Figure~\ref{fig:cross-correl-structures}, but a finer scale correspondence is also present within the structures themselves (Figure~\ref{fig:structure-brute}).
This is especially noticeable for the plasma density.
Of particular interest are four sub-structures, of duration between 0.1 and 0.3 h, detected at both spacecraft and numbered in Figure~\ref{fig:structure-brute}(a).
Such substructuring, with typical time scales $\sim$ 5-20 min, seem to be a common feature of solar wind density enhancements, and is most probably linked with their generation process at the Sun \citep{DiMatteo_2019_PDS_Helios, Kepko_2024_PDS_formed_at_Sun}.
We also note that the density structure ends in both observations (at $t \, \& \, t - \tau \sim 1.1 $~h) with a sharp decrease of similar short duration and magnitude (taking into account the spherical expansion).

With radial velocities between 200 and 300 km/s, the four density sub-structure durations translate to spatial sizes $L$ between $0.07$ and $0.5 \times 10^6$ km, so significantly below the estimated minimum distance $d_{MIN} \sim 7 \times 10^6$~km  ($\sim 2 \times 10^6$~km when including non-radial propagation, see Appendix~\ref{sec:Non-radial-propagation}). 
Since $L$ and $d_{MIN}$ are sizes estimations in two orthogonal directions (R and N), we conclude that, in order to be observed at both spacecraft, the finer observed structures could be elongated by at least a factor 4 in the N direction (about orthogonal to the ecliptic) as compared to the radial direction.  However, the difficulty is to estimate precisely enough the total amount of solar wind deflection in latitude with only velocity data measured at two locations of the trajectory (see Appendix~\ref{sec:Non-radial-propagation}).
Thus, the estimation of the extension ratio in the $R$ and $N$ directions of those finer density structures remains uncertain.

The physical parameters in Figure~\ref{fig:structure-brute} are shown as functions of time.
This way of comparing measurements may appear arbitrary, but, in the case of a structure varying along the radial direction, it appears as the most relevant one, as follows.
We show on Figure~\ref{fig:radial_stretching} a schematic of the radial distance $r$ as a function of time $t$ between two points representing the front (dashed line) and rear (dotted line) boundaries of the structure.
These two points have the same profile $r(t)$ up to a shift in time as represented in Figure~\ref{fig:radial_stretching}.
The spatial extent of the structure at a given time (green arrows) is increasing (or decreasing) with time because of the plasma acceleration (or deceleration).
However, for each given position, the time extent of the structure (red arrows) remains constant regardless of the acceleration profile.
Therefore, if the acceleration profile is the same along the structure, and neglecting the spacecraft radial speeds as compared to the solar wind's, the time difference between two parts of the structure should remain the same at the inner and outer spacecraft.
Such a behaviour already has been reported on a larger scale in \citet{Viall_2015_PDS} using white light images from STEREO's outer coronograph COR2. The authors observed a constant frequency (with a period $\sim 90$ min) of periodic density structures released near the tip of helmet streamers, despite their acceleration from $\sim 90$ to $\sim 180$ km/s within 2 to 15 solar radii.
Considering the observed structure to vary along the radial direction, this justifies the comparison in Figure~\ref{fig:structure-brute} of PSP and SolO data within the same time interval (only shifted by the transit time).

The studied plasma structure, shown in Figure~\ref{fig:structure-brute}, has indeed a similar duration of $\sim 1.5$ hours on the two spacecraft.
Taking into account the mean relative radial plasma velocity (with respect to the spacecraft) this corresponds to a radial spatial scale of $\sim 1.1 \times 10^6$~km and $\sim 1.8 \times 10^6$~km at PSP and SolO respectively, so about a factor $1.6$ of radial expansion between the two spacecraft.
This radial expansion is however much smaller than the expected longitudinal and latitudinal expansions ($\sim R_{SolO} / R_{PSP} = 12$).
Moreover, a radial stretching due to the solar wind acceleration would in general induce an additional decrease of $N_p$, as for a spherical expansion, the continuity equation implies
\begin{equation*}
    \partial_R(N_p V_p R^2) = 0 \Rightarrow N_p V_p \propto R^{-2}
\end{equation*}
with $\partial_R$ the partial derivative in the radial direction.
This in general would need to be taken into account when renormalizing the measurements in order to compare them.
We however found that in our case, this decrease was, coincidentally, compensated by a compression due to the formation of a stream interaction region (SIR).
Observations of solar wind density structures being swept up by SIRs already have been reported in previous studies \citep{Rouillard_2010_transients_a, Rouillard_2010_transients_b, Plotnikov_2016_Tracking_of_Corotating_Density_Structures, Kepko-Viall_2019_PDS_SIRs}, and a more thorough analysis of this particular case will be presented in a next article.


\section{Conclusions and Perspectives} \label{sec:conclusion}

We presented in this paper how we identified what we believe to be the same plasma (plasma line-up) passing through PSP ($\sim 0.075$~au) and SolO ($\sim 0.9$~au) after their radial alignment the 29/04/2021.
We began by modeling the plasma propagation considering a purely radial velocity, first with a constant speed using only PSP measurements (Section~\ref{sec:velocity_constante}), and then with a constant acceleration (Section~\ref{sec:acceleration_constante}) constrained by the measurements of both spacecraft.
This led to a first estimation of the time intervals corresponding to the plasma line-up.
A visual inspection paired with the use of a cross-correlation method finally allowed us to identify (Section~\ref{sec:plasma_identification}) the same density structure passing through both PSP and SolO.

Our main finding here is how well conserved the identified density structure is despite its $\sim 137$~h journey from PSP ($0.075$~au) to SolO ($0.9$~au).
We even are able to associate substructures with temporal scales of about $10-20$~min.  
There is also a (somewhat weaker) correspondence between the magnetic field magnitude at both spacecraft.
This may be simply a consequence of a total pressure equilibrium with the plasma pressure mainly modulated by its plasma density.

In general, identifying the same structure might not always be possible, even if the spacecraft are passing through the "same plasma".
Indeed, in order to recognize the same plasma structure, several hypotheses have to be fulfilled, as follows.
  \begin{itemize}
  \item The structure has to exist before reaching the inner spacecraft.
  \item Not only the structure should not be destroyed during its propagation, but it should also keep its identity enough as to be unambiguously recognizable between the two spacecraft.
  For example, \citet{Borovsky_2021_Structures_Non-Destroyed} presents some solar wind structures that are believed to not be destroyed under the effects of turbulence.
  \item The structure has to be large enough to pass through both spacecraft. Typically, mesoscale structures as described in \citet{Viall_2021_mesoscale_structures} (with radial sizes ranging from $\sim 5 \times 10^3$~km to $\sim 10^7$~km) usually have the right lengths for this, depending on the latitude difference between the two spacecraft. 
  \end{itemize}
Density enhancements generally fulfill these three conditions.
They are believed to be created by reconnection in the solar corona, and at least part of them are conserved during their propagation within the inner heliosphere.
This, is supported by elemental composition studies at 1 au \citep{Kepko_2016_PDS_composition_and_slow_wind_source, Kepko_2024_PDS_formed_at_Sun}.
Moreover, these structures can be large enough to pass through the two spacecraft as described in Section~\ref{sec:plasma_identification}.

Furthermore, the correspondence of the plasma structures at both spacecraft is best when the density measurements are compared as functions of time.
This implies that the structure mostly has radial gradients and was accelerated with a somewhat homogeneous velocity profile along it during the propagation (Figure~\ref{fig:radial_stretching}).
The discrepancies between the two recorded signals could moreover not only be due to the structure's evolution, but also to non-radial gradients.
We remind that the two spacecraft orbit with different azimuthal velocities and are at different latitudes around the studied intervals.
Moreover, in the present case, the solar wind also developed non-radial velocity components during its propagation (Figure \ref{fig:non-radial_speeds}).

The association of plasmas passing through two spacecraft solely using their \textit{in situ} measurements can be difficult since the provided data represent 1D temporal cuts through temporally evolving 3-dimensional plasmas. All the interpretations are therefore very dependent of the underlying assumptions made in order to compare these cuts.  Here, 3D MHD numerical simulations constrained with all possible data (\textit{in situ} and remote sensing) would be of great help to further confirm the association and to help understanding the physics involved.
It could also give a more physically relevant estimation of the solar wind propagation than a ballistic model.

The constant acceleration obtained in Section \ref{sec:acceleration_constante} is in agreement with the average velocity profiles derived in \citet{Maksimovic_2020_electrons} using Helios 1 \& 2 measurements as close as 0.3~au.
The more recent study of \citet{Dakeyo_2022_isopoly} using PSP and Helios 1 \& 2 data however shows that, in average, the slow solar wind tends to have a steeper acceleration closer to the Sun (within $[0.1, 0.3]$~au).
The modeling considering a constant acceleration might therefore no hold in a general case, and the fact that it is well relevant for our observations is probably due to the peculiarity of the identified intervals.

Lastly, the difference in latitude between the two spacecraft ($\sim 3$) imposes a minimum distance $d_{MIN} \simeq l_{\Delta \theta} \simeq 7 \times 10^6$~km (Section \ref{sec:propag_method}) limiting how close the plasma measured by the inner spacecraft can get to the outer one.
A more realistic estimation considering the latitudinal plasma deflection (Appendix \ref{sec:Non-radial-propagation}) gives $d_{MIN} \simeq 2 \times 10^6$~km.
In fact, this deflection might have played a major role by bringing the plasma closer to the outer spacecraft.
It is very likely that the structure would have missed SolO otherwise, making the plasma line-up identification not possible.
The density enhancement, as well as its substructures, should indeed be more extended and coherent in the north-south direction than this minimum distance $d_{MIN}$.

Finally, we only considered here the proton density $N_p$ and magnetic field's magnitude $B$.
However, other physical parameters also show interesting behaviours, requiring a deeper analysis which will be presented in the following study.
We will see that the identified intervals correspond to crossings of the heliospheric current and plasma sheets (HCS and HPS).
A SIR also developed during the plasma propagation, engulfing the HCS $\&$ HPS and sweeping up the density structure.


\begin{acknowledgements}
We are grateful for the helpful comments from Nicholeen Viall as a referee of this article.
\end{acknowledgements}

\bibliography{references}
\bibliographystyle{aa}

\begin{appendix}

\section{Consistency with a Non-radial Propagation}\label{sec:Non-radial-propagation}

\begin{figure}[t!] 
\resizebox{\hsize}{!}{\includegraphics{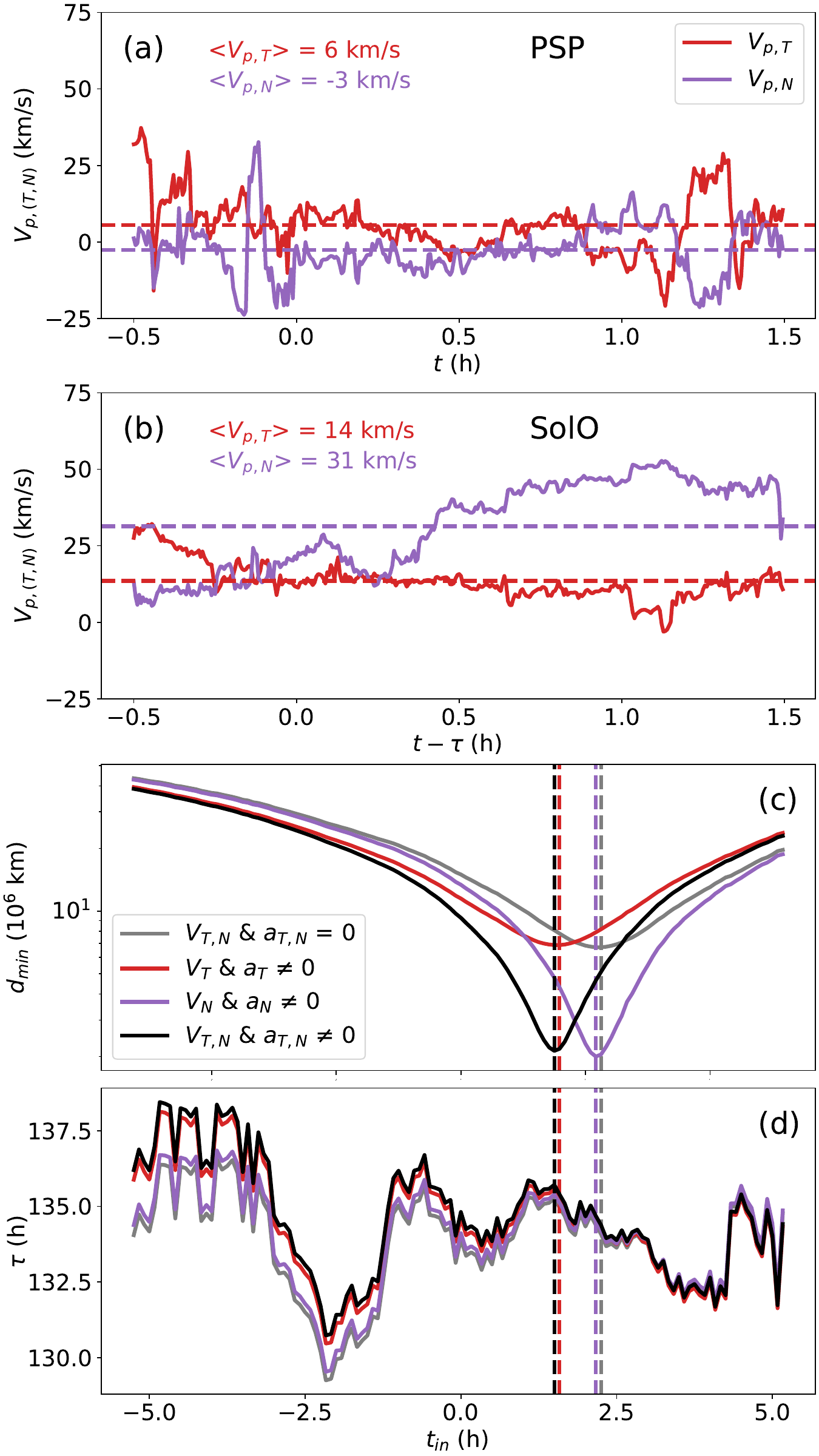}}
\caption{
Panels (a) and (b) show the $Vp,_T$ and $Vp,_N$ proton bulk speeds (red and purple curves respectively) measured at PSP and SolO for the same time intervals as in Figure \ref{fig:structure-brute}.
Panels (c) and (d) are the outcomes of the propagation method, as in Figure \ref{fig:outcomes_acceleration}, including some non-radial components.
Grey curves correspond to purely radial propagation, red and purple curves correspond respectively to propagation with non-zero T and N velocities and acceleration, and black ones to propagation with both T and N components.
For each propagation vector, the minimum distance $d_{MIN}$ is marked by vertical dashed lines of matching color.
The radial velocities and acceleration are the same as the ones in Figure \ref{fig:outcomes_acceleration}.
}
\label{fig:non-radial_speeds}
\end{figure}

In Section \ref{sec:propag_method}, the plasma line-up
is calculated considering the solar wind propagation to be purely radial.
This is not necessarily always justified as some phenomena can cause the solar wind to travel with non radial components.
We indeed show on Figure \ref{fig:non-radial_speeds}(a,b) that the protons bulk velocity has non-zero non-radial components around the identified plasma structures (especially for SolO).
We will see in the next study that this is due to the formation of a SIR.
We therefore investigate below the effect of these non-radial components on the modeled propagation (Section \ref{sec:acceleration_constante}).
The goal here is to see if, when considering these effects, the results are still consistent with the structure association done in Section \ref{sec:cross-correl}.

As in Section \ref{sec:acceleration_constante}, we suppose a constant acceleration during the plasma travel.
We consider here the RTN referential of the inner spacecraft (PSP) at $t_{in}$.
Since the equations used to model the propagation are written in vector form, we can extend the method by also scanning a relevant range of acceleration in the T and N directions.
The maximum acceleration values can be derived from Equation (\ref{eq:acceleration-vectorielle-finale}) following the same analysis leading to Equation (\ref{eq:a_max}).
This would require scanning a 3D space of parameters ($a_R, a_T, a_N$) to find the minimum of $d_{ min}$.
While numerically well achievable, we provide below an analytical approach to derive approximate estimates.

As a first approximation, and in order to highlight the different effects of the non-radial components, we consider independently the $T$ and $N$ velocity components.
We therefore compare 4 different cases
  \begin{eqnarray}
     \mathbf{V}(t) =& (V_R(t), 0, 0) \nonumber\\
     \mathbf{V}(t) =& (V_R(t), V_T(t), 0) \nonumber\\
     \mathbf{V}(t) =& (V_R(t), 0, V_N(t)) \nonumber\\
     \mathbf{V}(t) =& (V_R(t), V_T(t), V_N(t)) \nonumber
  \end{eqnarray}
with
  $$ 
    V_j(t) = V_{in,j} + (t - t_{in}) \, a_j
  $$ 
for $j = R, T, N$.

As a second approximation, the radial acceleration ($a_R$) and starting velocity ($V_{R,in}$) are kept as the ones obtained by minimizing $|\Delta V|$ in Section \ref{sec:acceleration_constante}, see Figure \ref{fig:outcomes_acceleration}.
We also consider $ \frac{|| \mathbf{V}_{out} + \mathbf{V}_{in} ||}{2 ||\Delta \mathbf{R}|| } \simeq \frac{V_{out,R} + V_{in,R}}{2 ||\Delta \mathbf{R}||}$, since the radial velocity is about one order of magnitude larger than the transverse velocity components.
This is equivalent to assuming that the propagation time of the plasma from the inner spacecraft to the nearest position of the outer spacecraft is mostly due to the radial component. The non-radial velocity components are large in the studied case (presence of a SIR, see next study) as compared to the usual solar wind. Still, as shown below, the inclusion of these non-radial components does not affect significantly $t_{in}$, $\tau$ and $t_{out}$ so that the plasma line-up is achieved with similar sets of data for both spacecraft.

Using Equation (\ref{eq:acceleration-vectorielle-finale}), the non radial acceleration components are written as
  \begin{equation}
    a_{j} = \frac{V_{out,R} + V_{in,R}}{2 ||\Delta \mathbf{R}||} \left( V_{out,j} - V_{in,j} \right) \label{eq:non-radial-acc}
  \end{equation}
with $j = T, N$.
Since the association between the measurements at PSP and SolO has already been done, we consider the non-radial speeds as $V_{(in, out), j} = \langle V_{p, (PSP, SolO), j} \rangle$ averaged over the time interval corresponding to the density structures.
$\langle V_{p, (PSP, SolO), j} \rangle$ are shown with horizontal dashed lines in Figure \ref{fig:non-radial_speeds}(a,b).
The primary goal here is to evaluate if a non-radial propagation challenges the correspondence between the two density structures.
This also makes the non-radial speeds and acceleration less dependent on $t_{in}$ and $t_{out}$, allowing an easier interpretation of their effects on the propagation.

We show $d_{min}$ and the corresponding $\tau$ as functions of $t_{in}$ for the different considered velocity vectors in Figure \ref{fig:non-radial_speeds}(c,d).
Grey curves correspond to a purely radial propagation, the same as shown in Figure \ref{fig:outcomes_acceleration}.
Red and purple curves correspond to radial propagations, to which we respectively added non-zero $T$ and $N$ components.
Black curves correspond to a propagation with the three $R$, $T$ and $N$ components.
Figure \ref{fig:non-radial_speeds}(c) shows that in this case, $\tau$ varies by at most $\sim 1$~h when considering a non-radial propagation, justifying a posteriori keeping the $a_R$ found in Section \ref{sec:acceleration_constante} and the use of Equation (\ref{eq:non-radial-acc}).

Modifying $V_{T}$ affects mostly the $t_{in}$ associated with $d_{ MIN}$.
Since the plasma has a finite propagation speed, a small longitude difference is needed between the two spacecraft in order for the outer one to intercept the closest possible plasma parcel. 
In this case, the plasma also propagates along the azimuthal direction, so, $\mathbf{V} (t)$ has now a small longitudinal angle and depending on this angle's sign, the longitude difference needed between the two spacecraft decreases (positive angle) or increases (negative angle).
This shifts the relationship between $d_{ min}$ and $t_{in}$ and almost does not change $\tau$ or $d_{ MIN}$ since the change in the travel distance and latitude difference, respectively, are small.
The shift is quite small due to  PSP's proximity to the Sun, making its longitude vary a lot as compared to SolO's one (see Figure \ref{fig:line-up-config_fused}), covering rapidly the change of longitude difference due to the inclusion of a finite $V_T$.

Modifying $V_{N}$ changes $d_{ min}$, so $d_{ MIN}$, see Figure \ref{fig:non-radial_speeds}(c).
As shown in Figure \ref{fig:line-up-config_fused}, PSP and SolO have a small latitudinal difference $\Delta \theta$.
As we already discussed in Section \ref{sec:propag_method}, this difference implies a distance $d_{MIN} \sim l_{\Delta \theta}$ when considering a purely radial propagation.
Adding a small positive $V_N$ reduces this distance because the plasma covers a part of the latitude difference during its propagation, lowering $d_{ MIN}$ to $\sim 2 \times 10^6$~km.
However, if the added $V_N$ is too big (such that at $t_{out}$ the plasma has a higher latitude than SolO), then, increasing $V_N$ would make $d_{ min}$ and $d_{ MIN}$ increase again (this is not the case with present data).

In summary, each velocity and acceleration component modify mostly one of the outcome's parameter: $\tau$ (and $t_{out}$) for $V_R$ \& $a_R$, $t_{in}(d_{min}=d_{MIN})$ for $V_T$ \& $a_T$ and $d_{MIN}$ for $V_N$ \& $a_N$.
The estimations for non-radial propagations are consistent with the structure's identification (Section \ref{sec:plasma_identification}) and do not change much the predicted time intervals for the plasma line-up.
In fact the inclusion a positive $V_N$ lowers $d_{MIN}$, reducing the estimated structures latitudinal extension.
The inclusion of a positive $V_T$ reduces the $t_{in}$ associated to $d_{MIN}$, bringing it closer to the observations, also reducing the needed structures longitudinal extension.
The inclusion of both $T$ and $N$ component implies a $d_{min}$ curve (black in Figure \ref{fig:non-radial_speeds}(c)) with, a close $d_{MIN}$ as when including $V_N \, \& \, a_N$ (purple Figure \ref{fig:non-radial_speeds} (c)), and at a nearby $t_{in}$ as when including $V_T \, \& \, a_T$ (red Figure \ref{fig:non-radial_speeds} (c)).
It is however difficult to check the exact extent of the deflection as the real propagation speed profile is certainly more complex than what we considered.
Moreover $V_T$ and $V_N$ are also varying along the structure, probably causing some distortion which would need to be taken into account.

\end{appendix}

\end{document}